\documentclass[11pt,english]{revtex4-1}
\usepackage[T1]{fontenc}
\usepackage[latin9]{inputenc}
\usepackage{geometry}
\usepackage{array}
\usepackage{enumitem}
\usepackage{amsmath}
\usepackage{amsthm}
\usepackage{amssymb}
\usepackage{graphicx}

\makeatletter

\providecommand{\tabularnewline}{\\}

  \theoremstyle{plain}
  \newtheorem{lem}{\protect\lemmaname}

\usepackage{color}

\makeatother

\usepackage{babel}
  \providecommand{\lemmaname}{Lemma}

\begin{document}
\global\long\def\sinc{\operatorname{sinc}}

\global\long\def\rank{\operatorname{rank}}

\global\long\def\spaan{\operatorname{span}}

\global\long\def\Var{\operatorname{Var}}

\global\long\def\Tr{\operatorname{Tr}}

\global\long\def\Re{\operatorname{Re}}

\global\long\def\Im{\operatorname{Im}}

\global\long\def\ket#1{|#1\rangle}

\global\long\def\bra#1{\langle#1|}

\global\long\def\braket#1#2{\langle#1|#2\rangle}

\global\long\def\Ket#1{\left|#1\right\rangle }

\global\long\def\Bra#1{\left\langle #1\right|}

\global\long\def\imag{\mathrm{i}}

\global\long\def\deriv{\mathrm{d}}

\global\long\def\bbone{\mbox{1\hspace{-0.85ex}{\large1}}}

\global\long\def\bbzero{\mbox{1\hspace{-0.85ex}{\large1}}}

\global\long\def\H{\operatorname{H}}

\global\long\def\op#1{\operatorname*\{#1\}}

\title{Quantum Process Identification: A Method for Characterizing Non-Markovian
Quantum Dynamics}

\author{Ryan S. Bennink}

\affiliation{Quantum Computing Institute, Oak Ridge National Laboratory}
\email{benninkrs@ornl.gov}

\author{Pavel Lougovski}

\affiliation{Quantum Computing Institute, Oak Ridge National Laboratory}
\begin{abstract}
Established methods for characterizing quantum information processes
do not capture non-Markovian (history-dependent) behaviors that occur
in real systems. These methods model a quantum process as a fixed
map on the state space of a predefined system of interest. Such a
map averages over the system's environment, which may retain some
effect of its past interactions with the system and thus have a history-dependent
influence on the system. Although the theory of non-Markovian quantum
dynamics is currently an active area of research, a systematic characterization
method based on a general representation of non-Markovian dynamics
has been lacking.

In this article we present a systematic method for experimentally
characterizing the dynamics of open quantum systems. Our method, which
we call \emph{quantum process identification} (QPI), is based on a
general theoretical framework which relates the (non-Markovian) evolution
of a system over an extended period of time to a time-local (Markovian)
process involving the system and an effective environment. In practical
terms, QPI uses time-resolved tomographic measurements of a quantum
system to construct a dynamical model with as many dynamical variables
as are necessary to reproduce the evolution of the system. Through
numerical simulations, we demonstrate that QPI can be used to characterize
qubit operations with non-Markovian errors arising from realistic
dynamics including control drift, coherent leakage, and coherent interaction
with material impurities.
\begin{table}[b]
\noindent This manuscript has been authored by UT-Battelle, LLC, under
contract DE-AC05-00OR22725 with the US Department of Energy (DOE).
The US government retains and the publisher, by accepting the article
for publication, acknowledges that the US government retains a nonexclusive,
paid-up, irrevocable, worldwide license to publish or reproduce the
published form of this manuscript, or allow others to do so, for US
government purposes. DOE will provide public access to these results
of federally sponsored research in accordance with the DOE Public
Access Plan (http://energy.gov/downloads/doe-public-access-plan).
\end{table}
\end{abstract}
\maketitle

\section{Introduction}

The experimental characterization of dynamical processes is fundamental
to physics. In the burgeoning field of quantum information science,
the characterization of quantum processes in a general, systematic
way has become particularly important. For example, in quantum key
distribution, characterization of the information-preserving properties
of the quantum channel between communicating parties is crucial to
establishing the security of the generated key \cite{Scarani2009}.
In the circuit model of quantum computing, a computation is represented
as a sequence of primitive quantum logic operations or ``gates'',
each of which is implemented via a controlled quantum process. Characterization
of these processes essential to assessing and improving their performance
\cite{Kelly2014,Dehollain2016,Sheldon2016,Blume-Kohout2017}. The
paradigmatic way of characterizing a process involving some quantum
system of interest is quantum process tomography (QPT) \cite{Nielsen2011}.
In QPT the process is tested many times using a variety of initial
states and final measurements that collectively span the state space
of the system. The process is then expressed as a \emph{quantum channel}\textemdash a
linear, completely positive, trace-preserving (CPTP) map on the system
state space\textemdash that can be estimated from the experimental
results.

A modern application of QPT (and the main context of this work) is
to construct models for the behavior of fabricated quantum computing
devices. Each qubit operation or ``gate'' the device is capable
of performing is characterized via QPT as a quantum channel on the
targeted qubits. The execution of an arbitrary gate sequence is then
modeled as a corresponding sequence of quantum channels on the device's
qubits. While newer and more robust characterization methods such
as gate set tomography (GST) \cite{Blume-Kohout2015,Dehollain2016,Blume-Kohout2017}
and randomized benchmarking (RB) \cite{Magesan2012-PRA,Magesan2012-PRL,Gambetta2012,Gaebler2012,Kimmel2014,Chasseur2015,Sheldon2016,Cross2016}
have largely replaced QPT, these methods continue the approach of
modeling each gate as a quantum channel involving only the targeted
qubits.

This nearly ubiquitous approach to quantum device characterization
is not valid, however, when the dynamics of interest are non-Markovian.
For the purposes of this work, a process is Markovian if the dynamical
map for any given time interval can be expressed as a sequence of
maps representing the dynamics over successive subintervals \cite{Buscemi2016}.
This property is called divisibility and may be expressed informally
as the property that the (statistical) state of the system at any
point in time is sufficient to determine its (statistical) state at
at any future time; there is no additional dependence on prior states
of the system or on any external degrees of freedom. By this definition
a sequence of quantum channels constitutes a Markovian process. While
the theory of non-Markovian quantum dynamics has received growing
attention in recent years \cite{Breuer2016,deVega2017}, an explicit,
systematic method for experimentally characterizing non-Markovian
dynamical processes has been lacking. Meanwhile, incoherent (Markovian)
noise processes in quantum information technologies have been reduced
to the point that non-Markovian errors are starting to becoming significant
\cite{Dehollain2016,Blume-Kohout2017}. 

In this paper we present a general way to empirically characterize
``black box'' quantum processes (i.e.~processes not involving intermediate
measurements), including non-Markovian dynamics of open quantum systems.
More specifically, we show how to construct a minimal dynamical model
of a quantum system solely from observations of that system at selected
times. We call this task \emph{quantum process identification} (QPI)
in analogy with the systems engineering task known as system identification
\cite{DeSchutter2000,Heij2007}. Operationally, QPI consists of time-resolved
process tomography followed by a numerical search for an appropriate
model. But unlike QPT and its modern replacements, QPI assumes almost
nothing about the system of interest, its environment, or their joint
dynamics. In particular it does not restrict analysis to a presupposed
system of interest, but identifies and characterizes as many degrees
of freedom as are needed to account for the observations. It assumes
no particular model, nor even quantum mechanics specifically. QPI
assumes only that (1) there exists a finite-dimensional probabilistic
model capable of predicting the observed behavior to desired accuracy,
and (2) experiments on the system are statistically independent. Furthermore,
QPI is fully self-contained; it requires no calibrated measurements
or physical references. In the context of quantum computing, QPI can
be used to better assess the quality of qubit operations and inform
the development of more effective control and error mitigation techniques.
Dynamical models produced by QPI may also be used to assess non-Markovianity
or quantumness in processes of interest; however, such questions are
beyond the scope of this work. We present QPI simply as a very general
method\textemdash free of the usual assumptions of Markovianity\textemdash to
obtain an accurate, empirically-derived mathematical model of a dynamical
system's observable behavior.

The development of QPI was motivated by studies addressing the limitations
of GST and RB for processes with non-Markovian aspects \cite{Wallman2014,Epstein2014,Ball2016,Blume-Kohout2017}
and by the emergence of ad-hoc adaptations of these methods to characterize
specific types of non-Markovian errors \cite{Chasseur2015,Wood2017}.
QPI provides a general, systematic solution to these challenges. QPI
is also closely related to the characterization of quantum channels
with memory \cite{Kretschmann2005,Rybar2015}. Those works address
a slightly different problem, however \footnote{In \cite{Kretschmann2005,Rybar2015}, each use of the channel involves
a newly prepared input state, while the environment's state persists
between successive uses. In the present work, the output of each process
(channel) becomes the input to the next; that is, both the system
state and environment state persist through successive repetitions
of the process.}, with QPI being more relevant to quantum computation and the latter
being more relevant to quantum communication. Other areas of study
directly related to QPI include the large subject of non-Markovian
quantum dynamics \cite{Breuer2016,deVega2017}, the limitations of
the quantum channel formalism \cite{Chitambar2015,Dominy2016,Pechukas1994},
and general frameworks for representing non-Markovian dynamics \cite{Cerrillo2014,Pollock2018}.
More broadly, QPI falls under the general topic of learning quantum
states and dynamics \cite{Mohseni2006,Aaronson2007,Bisio2010,daSilva2011,Huszar2012,Mahler2013,Lee2015,Howland2016,Monras2016,Haah2017,Carleo2017,Dumitrescu2016},
including dimension estimation \cite{Wehner2008,Gallego2010,Ahrens2012,Brunner2013,DAmbrosio2014}.
In particular, QPI can be viewed as inference of a quantum hidden
Markov model \cite{Monras2010,ONeill2012,Monras2016,Cholewa2017}.
Somewhat surprisingly, our method of \emph{quantum} process identification
is based on a 50-year-old method of model inference for \emph{classical}
discrete time linear systems, as will be explained.

In the remainder of this article we give a detailed description and
numerical demonstration of QPI. We begin in Section \ref{sec: gedanken}
with a thought exercise to develop the intuition underlying QPI and
illustrate several important principles. We then review in Section
\ref{sec: Ho-Kalman} a model inference technique from the field of
linear systems engineering that underlies QPI. The QPI protocol itself
is described in Section \ref{sec: QPI protocol}, followed in Section
\ref{sec: simulations} by simulation results which demonstrate the
effectiveness of QPI for representative non-Markovian quantum processes.
Lastly, in Section \ref{sec: discussion} we summarize our results
and identify possible directions of further development. Mathematical
and algorithmic details are provided in the Appendices.

\section{A Thought Exercise}

\label{sec: gedanken}
\begin{figure}
\begin{centering}
\includegraphics[scale=0.75]{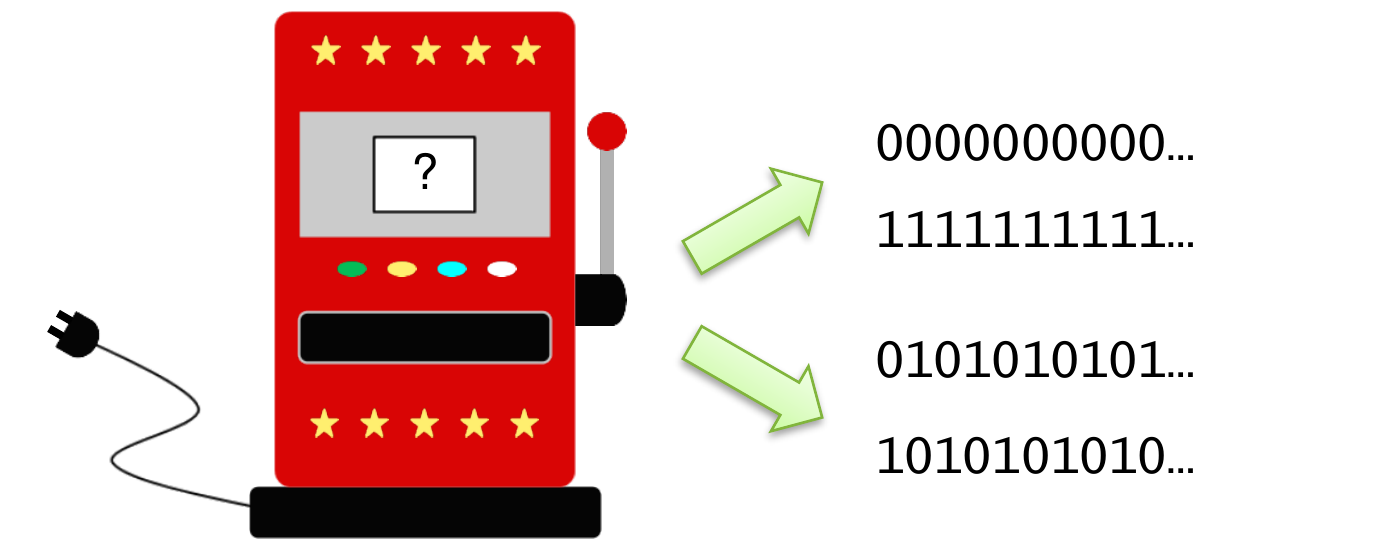}
\par\end{centering}
\caption{\label{fig: thought experiment}A fictional slot machine illustrates
how repeated observations of a system over time can be used to infer
the existence of, and to model, latent degrees of freedom responsible
for non-Markovian behavior. The slot machine produces a digit (0 or
1) each time its lever is pulled. A frequent player observes that
each time the slot machine is powered on it exhibits one of two distinct
behavior patterns: the output either alternates with each pull of
the lever or remains constant. This behavior cannot be explained by
any stochastic model involving the only the most recently output digit.
However, the behavior can be explained by positing the existence of
an internal bit that determines whether the output is constant or
alternating.}

\end{figure}

The means of detecting and characterizing non-Markovian
behavior can be understood through a simple example. Consider a slot
machine which produces a 0 or 1 each time its lever is pulled (Fig.\,\ref{fig: thought experiment}).
A frequent player ``Alice'' notices that each time the machine is
powered on, the first two digits produced are equally likely to be
$00$, $01$, $10,$ or $11$. If this were the extent of Alice's
observations, she might reasonably conclude that pulling the lever
simply outputs a random digit. That is, she would expect the probability
of observing a 1 would be 0.5, regardless of the value of previously
observed digits.

But suppose Alice observes longer output sequences and discovers that
each time the machine is powered on, one of two distinct behaviors
occurs: In half the cases, each pull of the lever repeats whatever
digit appeared first. In the remaining cases, the output alternates
between 0 and 1. Note that the value of the $(k+1)$th digit is not
predicted by the $k$th digit alone (the conditional probability is
$1/2$), but it \emph{is} predicted by the $(k-1)$th digit (they
are always the same). In other words, the next output of the machine
depends not only on the most recent output, but on prior outputs;
the output is non-Markovian. Alice can explain the observed behavior
by positing the existence of a ``control'' bit inside the machine
which determines whether the output is constant or alternating. The
control bit is set to a random value when the machine is powered on
and does not change. Such a model, involving unobserved degrees of
freedom that evolve according to some fixed (possibly stochastic)
rule, is known as a hidden Markov model \cite{Ghahramani2001}. Hidden
Markov models are widely used to model diverse phenomena in many different
fields.

Let us now imagine that the slot machine contains an internal counter
which causes the control bit to change state after $T$ pulls, thereby
switching the behavior from constant to alternating or vice versa.
If Alice only observes sequences of length $\le T$, she will see
nothing that indicates that the behavior of longer sequences is any
different; her model will be incomplete.

Several general principles may be gleaned from this example. The first
principle is that repeated interactions between a system and \emph{non-forgetful}
ancillary degrees of freedom can yield non-Markovian system behavior.
(Note that if the control bit was randomized at each time step, the
target bit would be randomized at each time step independent of its
previous value. In this case, there would be no need to track the
state of the control bit; a stochastic model of the target bit alone
would suffice.) The second principle is that the pattern\emph{ }of
system behavior over an extended period of time can be used to infer
the existence of, and determine the dynamics of, hidden degrees of
freedom. In this example the inference could be made simply by inspection,
but we will show how to do it in a systematic way. The third principle
is that since experiments can have only finite duration, it is not
possible to guarantee that one has observed enough behavior to obtain
a complete model, i.e. with enough degrees of freedom to accurately
predict all future behavior. But as we will show, it is possible to
guarantee that an experimental characterization yields a complete
model \emph{if} one has an upper bound for the effective dimension
of the process.

\section{Theory of Quantum Process Identification}

\subsection{Problem Formulation}

The problem we address is to how to experimentally characterize the
behavior of a quantum system of interest under some dynamical process,
without assuming anything about the system's size or composition
or the nature of the process. The process in question may be a repeatable
procedure $\mathcal{P}$ with definitive start and end, or it may
be a continuous, ongoing dynamic. In the latter case we let $\mathcal{P}$
denote evolution for some fixed time $\Delta\tau$, where $\Delta\tau$
becomes the time resolution at which the dynamics are resolved.

Ultimately, all that can be known about a system comes from one's
interactions with it. The kind of characterization we develop is relative
to, and in terms of, the available ways of interacting with the system.
In this work we consider only interactions in the form of ``experiments''
in which the system is prepared, made to evolve under $\mathcal{P}$,
and subsequently measured. (In section \ref{sec: discussion} we will
speculate how QPI may be extended to more complicated interactions.)
Let $\mathcal{I}$ be the set of ways the system can be prepared (initialized)
and let $\mathcal{M}$ be the set of ways it can be measured. An experiment
is a specified by a triple $(i,t,m)$ which specifies the initialization
$i\in\mathcal{I}$, followed by $t$ repetitions of $\mathcal{P}$,
followed by the measurement $m\in\mathcal{M}$. For notational convenience
we suppose that each measurement has only two possible outcomes, YES
or NO. (A measurement with $k$ possible outcomes can be cast as $k$
distinct YES/NO measurements.) Under these premises, the \emph{observable
behavior} of the system is the set of YES outcome probabilities for
all possible experiments. The YES probability for experiment $(i,t,m)$
will be denoted by $F_{i,m}^{(t)}$.

It is worth reiterating that the characterization of $\mathcal{P}$
produced by QPI is completely relative to the set of initializations
$\mathcal{I}$ and set of measurements $\mathcal{M}$ considered.
$\mathcal{I}$ and $\mathcal{M}$ need not be tomographically complete
for the system, but if they are not then the resulting characterization
may not address all aspects of the system. Furthermore, if an ``absolute''
characterization is desired, $\mathcal{I}$ and $\mathcal{M}$ should
include the procedures that define the reference frame. As a side
result, QPI also yields a characterization of the initializations
and measurements relative to one another.

For our purposes, characterization amounts to constructing a mathematical
model of the process $\mathcal{P}$, initializations $\mathcal{I}$,
and measurements $\mathcal{M}$. We opt not to use the traditional
Hilbert-space formulation of quantum mechanics, but instead the more
general framework known as generalized probabilistic theory (GPT)
\cite{Hardy2001,Barrett2007,Masanes2011}. GPT is a framework for
a class of physical theories that includes quantum mechanics and classical
mechanics as special cases. For us GPT has two appealing features:
First, physical phenomena are described in operational terms. Rather
than expressing phenomena in terms of complex operators that are not
directly observable, GPT expresses phenomena in terms of the outcome
probabilities of measurements that could be performed. This matches
the task of experimental characterization extremely well. Secondly,
the GPT formalism does not distinguish between quantum and classical
behaviors; it treats all behaviors in a uniform way. This is useful
because although the system of interest and the environment are nominally
quantum, their behavior may have some purely classical aspects. GPT
allows one to construct the most economical model of a process without
choosing it to be quantum or classical. On the other hand, some symmetries
required by quantum mechanics (such as complete positivity) are less
conveniently expressed in GPT. We emphasize that QPI is not tied to
the GPT formalism; if desired, QPI could be formulated entirely in
terms of the traditional Hilbert space formalism.

One approach to representing non-Markovian dynamics is to construct
dynamical maps that are not linear and/or not CPTP \cite{Dominy2016}.
Another strategy is to go beyond time-local maps and use structures
that explicitly encode temporal correlations, such as spectral densities
\cite{Addis2014,Guarnieri2014}, convolution-like maps \cite{Cerrillo2014},
or process tensors \cite{Pollock2018}. QPI takes a different approach:
the system's temporal evolution (whether Markovian not) is modeled
by a fixed \emph{Markovian} process involving both the system of interest
and a sufficiently large abstract environment. (This is analogous
to Stinespring dilation of a CPTP channel \cite{Stinespring1955,Wood2015}.)
That is, an extended Markovian model is formulated to predict the
system's observable (possibly non-Markovian) behavior. In technical
terms, the central premise of QPI is that \emph{the system's observable
behavior can be modeled to desired accuracy by a sufficiently large
finite-dimensional, time-independent, time-local dynamical model}\footnote{Time-independent means the dynamical equations have no explicit dependence
on time. Time-local means that all quantities in the dynamical equation
are evaluated at the same time. }. Significantly, such a model can be inferred from measurements that
nominally involve the system alone. This is important because the
experimentalist often does not have access to, or even knowledge of,
all the pertinent parts of the environment. By inferring a model on
an extended state space, QPI implicitly characterizes the involvement
of pertinent environmental degrees of freedom (though these degrees
of freedom are abstract and might not be readily associated with particular
physical degrees of freedom).

The only other assumption of QPI is that \emph{each experiment is
statistically independent. }One way of satisfying this second assumption
is to use strong initializations that erase (as far as can be detected)
any correlations between the system and the environment due to previous
experiments. For example, the initialization procedure may include
waiting a time much longer than the plausible memory lifetime of the
environment. Note, however, that we allow the initialization procedure
itself to correlate the system and environment. A weaker way of satisfying
the independence assumption is to randomize the order of the experiments
so that any residual correlations between the system and environment
at the start of each experiment are sampled fairly. In this case,
QPI infers a distribution of correlated initial states.

In QPI, a model of dimension $d$ consists of a state vector $s_{i}\in\mathbb{R}^{d}$
for each initialization procedure $i\in\mathcal{I}$, a property vector
\footnote{A measurement with a YES/NO outcome space may be regarded as a determination
of whether or not a system has a particular property.} $p_{m}\in\mathbb{R}^{d}$ for each measurement procedure $m\in\mathcal{M}$,
and a transfer matrix $T\in\mathbb{R}^{d\times d}$ for the process
$\mathcal{P}$. Then the YES probability of experiment $(i,t,m)$
is 
\begin{align}
F_{i,m}^{(t)} & =s_{i}T^{t}p_{m}.
\end{align}
(We have adopted the convention in which states are row vectors and
operators are applied on the right.) In terms of the matrices $S=[\begin{array}{ccc}
s_{1}; & s_{2}; & \ldots\end{array}]$ and $P=[\begin{array}{ccc}
p_{1} & p_{2} & \ldots\end{array}]$, we have
\begin{align}
F^{(t)} & =ST^{t}P.\label{eq: X =00003D STP}
\end{align}
The goal of QPI is to determine a value of $d$ and matrices $S$,
$P$, and $T$ that accurately predict $F^{(t)}$ for all $t=0,1,2,\ldots$,
i.e. the observable behavior of the system as defined above. Here
$S$ and $P$ are a characterization of the state preparations and
measurements, respectively, while $T$ characterizes the process $\mathcal{P}$
itself. We note that these matrices can be determined only up to a
similarity transformation, since for any invertible matrix $G$ the
mapping $S\rightarrow SG$, $P\rightarrow G^{-1}P$, $T\rightarrow G^{-1}TG$
leaves all probabilities unchanged. This indeterminacy of representation
is known as ``gauge indeterminacy''. While gauge indeterminacy does
lead to a minor theoretical conundrum regarding the assessment of
process fidelity \cite{Proctor2017}, it has (by definition) no experimental
consequence and thus is not a limitation of QPI.

In the case of a continuous-time process, QPI can be used to obtain
an effective master equation. Let $\rho(\tau)\in\mathbb{R}^{d}$ denote
the extended system-environment state at time $\tau$ with $F^{(\tau)}\equiv\rho(\tau)P$.
From (\ref{eq: X =00003D STP}) we have $\rho(t\Delta\tau)=ST^{t}=\rho((t-1)\Delta\tau)T$
where $\Delta\tau$ is the chosen time resolution. For sufficiently
small $\Delta\tau$, $T\approx1+L\Delta\tau$ for some matrix $L$.
Then $\rho$ approximately obeys the master equation
\begin{align}
\frac{\deriv\rho(\tau)}{\deriv\tau} & =\rho(\tau)L.
\end{align}

\subsection{The Ho-Kalman Method}

\label{sec: Ho-Kalman}A key to solving the problem of characterizing
non-Markovian quantum processes is a technique that was devised half
a century ago in the context of discrete-time linear systems and is
now a standard topic in introductory texts on systems engineering
\cite{DeSchutter2000,Heij2007}. Consider a classical input-output
system whose output $x\in\mathbb{R}^{n}$ at each time $t\in\mathbb{Z}$
is a linear function of its input $u\in\mathbb{R}^{m}$ at the previous
$d$ times. Ho and Kalman \cite{Ho1966} showed that the system can
be described by a $d$-dimensional state model of the form
\begin{gather}
\begin{aligned}s(t+1) & =u(t)A+s(t)B\\
x(t) & =s(t)C
\end{aligned}
\label{eq: Ho-Kalman state model}
\end{gather}
where $s\in\mathbb{R}^{d}$, $A\in\mathbb{R}^{m\times d}$, $B\in\mathbb{R}^{d\times d}$,
and $C\in\mathbb{R}^{d\times n}$. Furthermore, they showed how to
obtain the coefficients $A,B,C$ from the observable quantities $X(1),\ldots,X(2d)$
where $X_{i,j}(t)$ is the value of the output $x_{j}(t)$ in response
to a unit impulse on the $i$th input ($u_{i}$) at time 0. In this
case $X(t+1)=AB^{t}C$. Let 
\begin{align}
H & =\left[\begin{array}{ccc}
X(1) & \cdots & X(d)\\
\vdots &  & \vdots\\
X(d) & \cdots & X(2d-1)
\end{array}\right]
\end{align}
and 
\begin{align}
H^{\prime} & =\left[\begin{array}{ccc}
X(2) & \cdots & X(d+1)\\
\vdots &  & \vdots\\
X(d+1) & \cdots & X(2d)
\end{array}\right].
\end{align}
The key fact here is that $H$ and $H^{\prime}$ have related rank-$d$
factorizations, $H=LR$ and $H^{\prime}=LBR$ where $L=[A;\thinspace AB;\thinspace\cdots;\thinspace AB^{d-1}]$
and $R=[C\thinspace BC\thinspace\cdots B^{d-1}C]$. The Ho-Kalman
result is that a model of the form (\ref{eq: Ho-Kalman state model})
can be obtained by finding any rank-$d$ factorization $H=LR$, taking
the first $m$ rows of $L$ for $A$, the first $n$ columns of $R$
for $C$, and taking $B=L^{+}H^{\prime}R^{+}$ where $^{+}$ denotes
the Moore-Penrose pseudo-inverse.

The relevance of this result to QPI becomes clear upon noting that
$X(t+1)=AB^{t}C$ has the exactly same form as eq.~(\ref{eq: X =00003D STP}).
From the perspective of process tomography, $A$ describes the preparable
states and $C$ describes the measurable properties. When $d>\min(m,n)$
these do not span the state space of the system, yet a full reconstruction
of the process is still possible. That is, observations of the system
over a sufficiently long period of time can yield complete information
about the system, even when the preparable states and measurable properties
are not intrinsically complete.\emph{ }The critical implication of
this result for QPI is that \emph{the properties of a system of interest
over time are sufficient to reconstruct an accurate model of all relevant
dynamics, including the dynamics of any external degrees of freedom
with which the system interacts.}

Mathematically, the construction of a $d$-dimensional model from
$H$ is possible because $H$ has rank $d$. What if $A$ and $C$
are already tomographically complete? In that case it ought to be
sufficient to use just $X(1)=AC$ for $H$ and $X(2)=ABC$ for $H^{\prime}$,
as in QPT. In this case, the textbook Ho-Kalman method asks for more
data than is actually necessary. The following Lemma, proved in Appendix
\ref{sec: proof of Lemma}, states that the number of time steps needed
to ensure a complete characterization is determined by the number
of latent (unmeasured) degrees of freedom only, not the total number
of degrees of freedom:
\begin{lem}
Let $X(t)$ be the impulse response function for a linear, discrete-time,
time-shift invariant system. Let
\begin{align}
H^{(t;k)} & \equiv\left[\begin{array}{cccc}
X(t) & X(t+1) & \cdots & X(t+k)\\
X(t+1) &  &  & \vdots\\
\vdots &  &  & X(t+2k-1)\\
X(t+k) & \cdots & X(t+2k-1) & X(t+2k)
\end{array}\right]
\end{align}
and let $r=\max_{t}\rank X^{(t)}$. If the system has dimension $d$,
then $\rank H^{(1;l)}=d$ for $l\ge d-r$. Furthermore a minimal ($d$-dimensional)
model of the system can be constructed from $X^{(1)},\ldots,X^{(2l+2)}$.
In particular, if $LR$ is any rank factorization of $H^{(1;l)}$,
then $X(t+1)=AB^{t}C$ where $A$ is the first $m$ rows of $L$,
$C$ is the first $n$ columns of $R$, and $B=L^{+}H^{(2;l)}R^{+}$.
Observation of the system at fewer than $2l+2$ times is generally
insufficient to construct a complete model, as is observation of the
system at non-consecutive times.
\end{lem}
The Ho-Kalman method is the basis of our approach to QPI. However,
there are two important differences between a (classical) linear input-output
system and a quantum process that require consideration:
\begin{itemize}
\item In the Ho-Kalman problem $X_{i,j}(t)$ is (in principle) a directly
observable quantity, whereas in QPI the measurable quantity $F_{i,m}^{(t)}$
is a probability. One consequence is that in QPI, many repetitions
of an experiment are needed to obtain each value $F_{i,m}^{(t)}$.
Even then, one only obtains an estimate $\tilde{F}_{i,m}^{(t)}$ whose
precision is limited by the number of repetitions of the experiment.
There is a large body of work on the extension of the Ho-Kalman approach
to noisy data, but it considers additive or multiplicative Gaussian
noise and is not applicable to QPI, where $\tilde{F}_{i,m}^{(t)}$
has a binomial distribution whose width depends on the (unknown) value
of $F_{i,m}^{(t)}$.
\item The observables in the Ho-Kalman method are all mutually compatible.
In principle, the values $X_{i,j}(t)$ for all $j$ and all $t\ge1$
can be obtained in a single experiment with initial state $i$. In
contrast, measuring a quantum system generally disturbs it (an aspect
of the uncertainty principle). Thus in QPI, not all measurable properties
of the system can be measured at the same time, nor is the evolution
of the system after a measurement the same as if the system had not
been measured. Each $F_{i,m}^{(t)}$ must be estimated from a distinct
set of experiments, in which the system evolves unobserved for $t$
process repetitions and is then measured.
\end{itemize}
In the next section we present a method for QPI based on an extension
of the Ho-Kalman approach to the quantum domain.

\section{The QPI Protocol}

\label{sec: QPI protocol}

\subsection{The Experimental Protocol}

\label{sec: experimental protocol}
\begin{figure}
\begin{centering}
\includegraphics{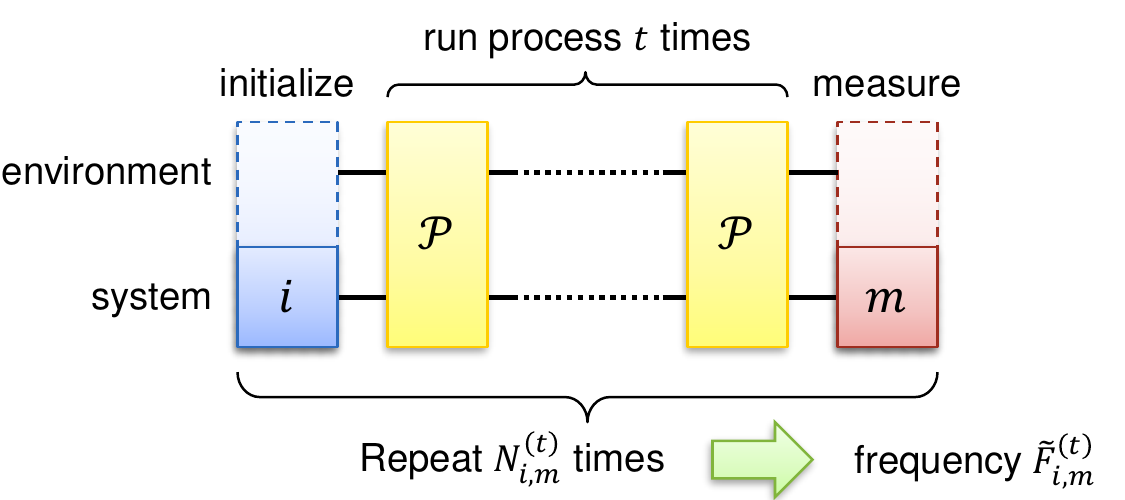}
\par\end{centering}
\caption{\label{fig: experimental procedure}The experimental procedure for
quantum process identification. An experiment consists of an initialization
$i$, $t$ repetitions of the process, and a final YES/NO measurement
$m$. Initializations and measurements nominally concern the system
but are considered to involve the environment as well. Each experiment
is repeated many times, yielding a frequency of YES outcomes. For
the most complete and accurate results, all available initializations
and measurements should be used, and the set of $t$ values should
consist of widely spaced ``flights'' of consecutive values (see
Section \ref{sec: experimental protocol}). }

\end{figure}

QPI is accomplished by performing
a series of experiments, each repeated many times (Fig.~\ref{fig: experimental procedure}).
Each experiment is designated by a triple $(i,t,m)$ which designates
the $i$th way of initializing the system, followed by $t$ repetitions
of the process to be characterized, followed by the $m$th way of
measuring the system. For each experiment $(i,t,m)$ one records $N_{i,m}^{(t)}$,
the number of times the experiment was performed, and $Y_{i,m}^{(t)}$,
the number of times the YES outcome was observed. The frequency $\tilde{F}_{i,m}^{(t)}\equiv Y_{i,m}^{(t)}/N_{i,m}^{(t)}$
is the experimental estimate of $F_{i,m}^{(t)}$.

The set of experiments performed must be chosen with some care in
order for QPI to be effective. If the experimental probabilities could
be determined perfectly, then by Lemma 1 $F^{(0)},\ldots,F^{(2l+1)}$
would be sufficient to characterize any process of dimension $d\le l+\rank F^{(0)}$.
(Note that $\rank F^{(0)}$ is just the dimension of the space spanned
by the initial states and measurements. If these are tomographically
complete for the system of interest, $\rank F^{(0)}$ is just the
system dimension.) However, some of the degrees of freedom may be
slow to manifest in the sense that it takes many repetitions of the
process for their impact on observable properties to become significant.
In such cases the error-free matrix $H^{(0;l)}$ is ill-conditioned
and even a small amount of statistical error in the data yields an
extremely poor characterization. For this reason it is advantageous
to perform experiments covering a wide range of $t$ values. As a
general rule, an experiment with $2t$ process repetitions is twice
as sensitive to the process parameters $T$ as an experiment with
$t$ process repetitions \cite{Blume-Kohout2015}.

To efficiently cover a wide range of $t$ values while ensuring the
data is informationally complete, experiments are chosen to form ``flights''
(Fig.~\ref{fig: data structures}a). A flight is a set of experiments
with $2l+1$ consecutive $t$ values and all values of $i,m$ for
each $t$. Each flight contains enough information to determine the
process on a space of dimension up to $l+\rank F^{(0)}$. Using many
flights with widely separated ranges of $t$ provides increased sensitivity
to the elements of the process matrix $T$, while presenting opportunities
to detect and characterize even more latent variables. To preserve
the factorable structure of $H$ required by the Ho-Kalman method,
the flights are chosen to have base $t$ values that form a biexponential
set $\{\varrho_{a}+\varrho_{b}:~a\in\{0,1,2,\ldots,\bar{a}\},\thinspace b\in\{0,1,2,\ldots,\bar{b}\}\}$
where $\bar{a},\bar{b}$ determine the number of flights and
\begin{align}
\varrho_{i} & =\begin{cases}
0 & i=0\\
2^{i-1} & i>0
\end{cases}
\end{align}
(see Fig.~\ref{fig: data structures}b). Putting this all together,
the experiments performed are those with $t$ values in the set
\begin{align}
\mathcal{T} & =\{\varrho_{a}+\varrho_{b}+k:~a\in\{0,\ldots,\bar{a}\},\thinspace b\in\{0,\ldots,\bar{b}\},\thinspace k\in\{0,\ldots,2l\}\}
\end{align}
and with all values of $i,m$. We found that this pattern of experiments
is generally necessary to obtain accurate models, especially for systems
in which the non-Markovian aspects manifest only over long timescales.
Notably, the characterization accuracy grows as $\max\mathcal{T}$,
whereas the number of experiments to be performed grows asymptotically
as only $(\log\max\mathcal{T})^{2}$.

\subsection{Model Inference Procedure}

\label{sec: model inference}Once the experiments are performed, the
data is analyzed to infer a model of the process $\mathcal{P}$, the
initial states, and the measurements. Our basic approach is to seek
the model with the highest likelihood for the given observations.
This is a challenging task for two reasons: (1) The size of the model
(number of dimensions) is unknown. (2) The likelihood is extremely
nonlinear in $T$: $T$ appears with powers up to $\max\mathcal{T}$,
which in practice can be on the order of $10^{3}$ or more. This creates
an extremely unforgiving optimization landscape with many local (and
very suboptimal) maxima.

Regarding the first challenge, a number of techniques have been devised
to compare the likelihood of models with different numbers of parameters
(e.g. the Akaike information criterion \cite{Akaike1974}) or to construct
models whose complexity automatically scales with the complexity of
the data (such as Dirichlet processes \cite{Buntine2010}). We use
a simple but effective technique to first estimate the dimensionality
of the process in question, which then determines the set of parameters
to be estimated. The second challenge is essentially the same as that
faced in gate set tomography. Building upon the progressive fitting
approach described in \cite{Blume-Kohout2015}, we devised a 4-stage
model inference procedure, carefully honed over the course of our
studies to reliably find concise, accurate process models. An overview
of our procedure is given below; additional details are given in Appendices
\ref{sec: dimension estimation}-\ref{sec: final estimate}.

\begin{figure}
\begin{centering}
\begin{tabular}{cc}
\includegraphics[scale=0.8]{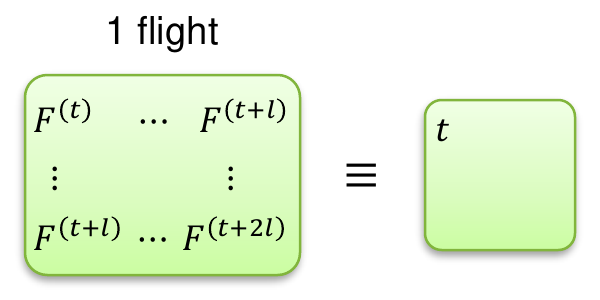} & \tabularnewline
(a) & \tabularnewline
\end{tabular}
\par\end{centering}
\begin{centering}
\begin{tabular*}{6.5in}{@{\extracolsep{\fill}}>{\centering}m{3.25in}>{\centering}m{3.25in}}
\includegraphics[scale=0.75]{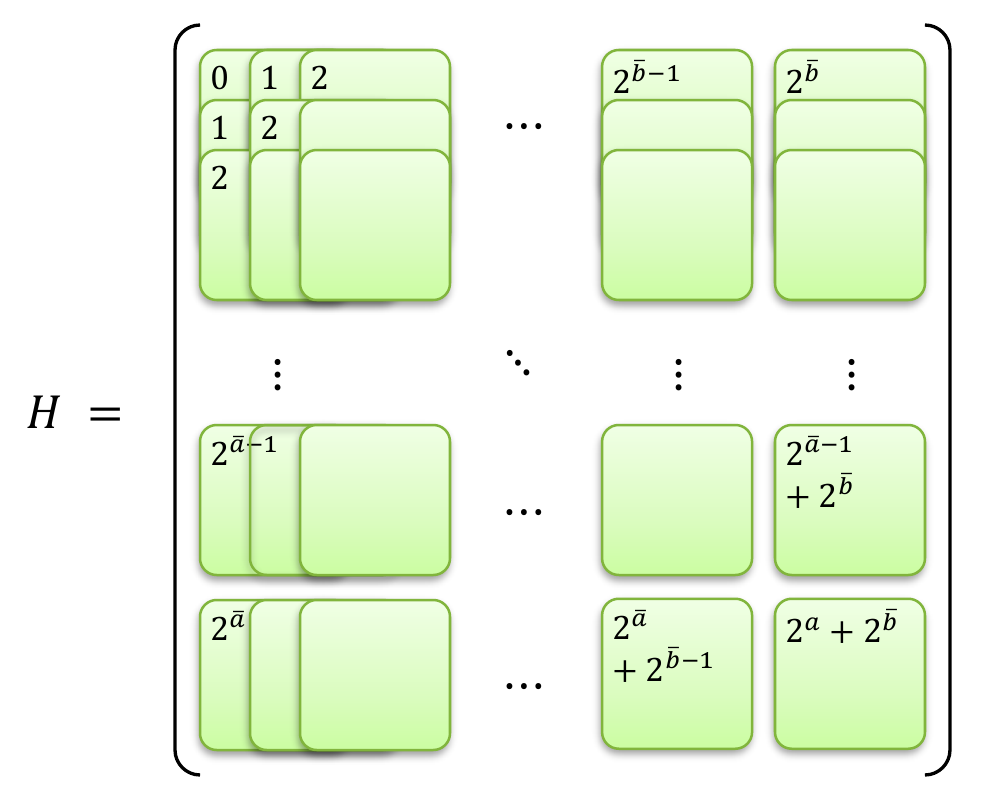} & \includegraphics[scale=0.75]{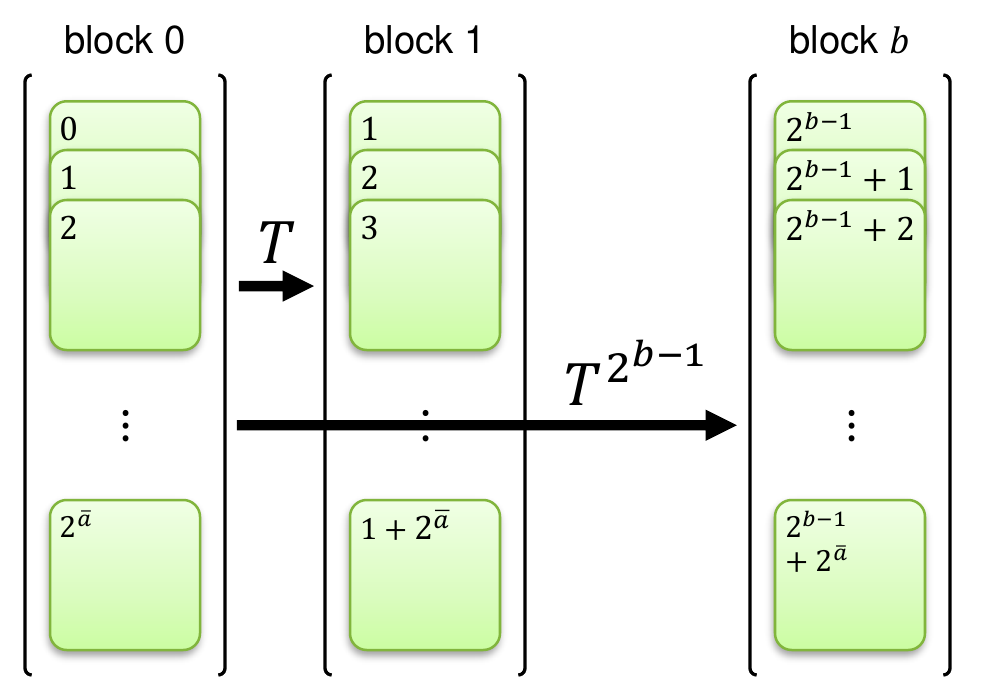}\tabularnewline
(b) & (c)\tabularnewline
\end{tabular*}
\par\end{centering}
\caption{\label{fig: data structures}Arrangement of experimental data for
analysis. (a) A flight is set of experiments with $2l+1$ consecutive
values of process repetitions $t$, and is denoted by its lowest $t$
value. (b) For dimension estimation and initial model estimation,
flights are arranged to form a large Ho-Kalman matrix. (c) Nonlinear
model fitting is performed by grouping flights into blocks. The data
in block $b$ is related to the data in block 0 by a factor of $T^{2^{b-1}}$,
where $T$ is to be determined. }

\end{figure}
\begin{enumerate}
\item \textbf{Estimate the Model Dimension}. First, the data is arranged
into a matrix $\tilde{H}$ which is the experimental estimate of the
Ho-Kalman matrix $H$ (Fig. \ref{fig: data structures}b). The effective
dimension $d$ is estimated as the number of statistically significant
non-zero singular values of $\tilde{H}$ \cite{Zeiger1974}, where
significance is assessed by a type of chi-squared test. In brief,
the uncertainties in the experimental data are estimated and propagated
to yield uncertainties in the singular values of $\tilde{H}$. This
yields for each $k=0,1,2,\ldots$ a threshold for the sum of squares
of the $k$ smallest singular values. If for some $k$ the sum falls
below the threshold, those singular values are discounted (deemed
insignificant). The dimension $d$ is taken to be the number of singular
values that remain after the largest subset of singular values has
been discounted. We note that this procedure does not determine the
inherent dimension of a process so much as the dimension that is warranted
by the data.
\item \textbf{Obtain an Initial Model}. This stage is a variation of the
Ho-Kalman method, modified to account for uncertainty in the experimental
data. Let $\tilde{H}^{\prime}$ be the ``time shifted'' version
of $\tilde{H}$, i.e. the matrix obtained by increasing the $t$ value
of each element of $\tilde{H}$ by 1. First, the variance of each
experimental quantity $\tilde{F}_{i,m}^{(t)}$ is estimated. Then
a rank-$d$ factorization $\tilde{H}\approx LR$ is found by minimizing
$\sum_{i,m}W_{i,m}(LR-\tilde{H})_{i,m}^{2}$ where $W_{i,m}$ is the
inverse variance of $\tilde{H}_{i,m}$. The process matrix is estimated
as the matrix $T$ that minimizes $\sum_{i,m}W_{i,m}^{\prime}(LTR-\tilde{H}^{\prime})_{i,m}^{2}$
where $W_{i,m}^{\prime}$ is the inverse variance of $\tilde{H}_{i,m}^{\prime}$. 
\item \textbf{Obtain an Improved Model.} Although the previous stage uses
all the data, it uses only the fact that $\tilde{H}$ and $\tilde{H}^{\prime}$
are related by a (single) factor of $T$. It does not use the fact
that data from different flights should be related by higher powers
of $T$. To incorporate this relationship into the model estimate,
the data is organized into groups of flights or ``blocks'' with
exponentially increasing $t$ values (Fig.\,\ref{fig: data structures}c).
Starting with the matrices $L,T,R$ obtained in stage 2, a search
is performed to find matrices $A,T,B$ that best fit the data in all
the blocks, from which $S$ and $P$ are also extracted. To ensure
that the search finds a good solution and does not get stuck in a
local extremum, the search is performed progressively, starting with
the lowest-$t$ blocks and gradually incorporating data from blocks
with higher $t$ values. The low-$t$ blocks yield coarse estimates
of $T$ which are gradually refined using the data in higher-$t$
blocks.
\item \textbf{Obtain a physically valid model estimate.} The model produced
by stage 3 is usually fairly accurate, but tends to make slightly
unphysical predictions (i.e., probabilities outside the interval $[0,1]$).
In this last stage, the model likelihood is maximized with the addition
of penalty terms to encourage the validity of predicted probabilities.
As in stage 3, the log-likelihood is approximated by a weighted sum
of squared errors; however, this time the weights are not based on
the (fixed) experimental frequencies but are based on the predicted
probabilities. Also, in this step the data is not arranged into Ho-Kalman
matrices; the model (\ref{eq: X =00003D STP}) is fit directly to
each experiment.
\end{enumerate}
In our simulation studies, the model inference procedure described
above proved to be both fast and reliable. In over 99\% of the thousands
of simulations we performed, it found a high-likelihood model without
any intervention. In the few cases in which it did not, making minor
adjustments to optimization parameters in stage 3 or 4 (e.g. penalty
weights, convergence thresholds) invariably led to success. We found
that all four stages were usually necessary to obtain accurate models.
On a standard desktop computer, the time to complete all four stages
ranged from just a few seconds for 7 dimensional processes to a few
minutes for a 19-dimensional process.

\section{Simulations}

\label{sec: simulations}To demonstrate the effectiveness of QPI,
we simulated the characterization of several different non-Markovian
processes relevant to quantum computing. Each process consists of
a simple unitary operation on a qubit with a weak, physically-motivated
non-Markovian error. In each case, the system of interest is the qubit
and the non-Markovian behavior of the qubit is generated from a Markovian
model of the qubit interacting with some external degree(s) of freedom.
For these studies the error processes were chosen to be not only weak
but also slow to reveal their non-Markovian nature, in order to present
a strong characterization challenge.

For each process, we first simulated the evolution of the quantum
state for several different initial states and predicted the time-dependent
outcome probabilities for the standard tomographic measurements. The
``standard tomographic measurements'' are the measurements of the
dimensionless Pauli operators $\sigma_{x},\sigma_{y},\sigma_{z}$.
These operators have eigenvalues $+1$ (``YES'') and $-1$ (``
NO'') with corresponding eigenstates denoted $\ket{\pm}_{x}$,$\ket{\pm}_{y}$,$\ket{\pm}_{z}$,
respectively. For simplicity, state preparation and measurement were
assumed to be ideal. Then, we simulated the collection of experimental
data by choosing a random number of YES outcomes for each measurement
from the computed probability distribution. The simulated data for
each experiment was then passed to our data analysis code, which estimated
the dimension of the process and found the most likely model using
the procedure outlined above. The error of the characterization was
calculated by using the inferred model to predict the time-dependent
qubit state and comparing the predicted state to the true (simulated)
state. The error at each time step was averaged over all initial states
and over 500 independent simulated characterizations (200 for the
system described in section \ref{sec: 31P study}).

In each plot below, black dots show the times of the (simulated) measurements
and their average error, where error is defined as the trace distance
$\frac{1}{2}\Tr\left|\rho(t)-\rho_{\text{est}}(t)\right|$ between
the true state $\rho(t)$ and the state $\rho_{\text{est}}(t)$ estimated
from the measurements at time $t$ only. The blue line is the QPI
error, defined as the average trace distance between the true state
$\rho(t)$ and the state $\rho_{\text{QPI}}(t)$ predicted by QPI,
which is based on all the simulated measurements. For comparison,
the dotted black line is the corresponding error for \emph{perfect}
quantum process tomography, obtained using the exact measurement probabilities
at $t=0$ and $t=1$ only.

\subsection{Qubit Rotation with Slowly Varying Systematic Error}

In quantum computing technologies, operations on qubits are typically
accomplished by applying control pulses that temporarily modify the
Hamiltonian to induce a unitary rotation of one or more qubits. In
many implementations the amount of rotation is proportional to the
pulse energy. A miscalibration of the pulse intensity or duration
results in over- or under-rotation of the qubits, which generally
contributes to computational error \cite{Barnes2017}.

In this example the process of interest is an intended $\pi$ rotation
of a qubit about the $y$ axis, i.e. a bit flip. We suppose however
that the intensity of the control pulse drifts sinusoidally over time,
such that the actual angle of rotation produced by the pulse at time
$t$ (where each pulse takes one unit of time) is
\begin{align}
\theta_{t} & =\pi+\epsilon\sin\Omega t.
\end{align}
We choose $\epsilon=0.01$ and $\Omega=0.02$, which yields a small,
slow oscillation of the qubit about the intended state in the $x$-$z$
plane. This oscillation cannot be described by any fixed Markovian
process involving only the qubit. In fact, the error can be described
as phase modulation, a process which has an infinite number of sidebands.
In other words, this process technically has an infinite dimension.
However, relatively few dimensions are needed to accurately reproduce
typical experimental data.

\begin{figure}
\begin{centering}
\includegraphics{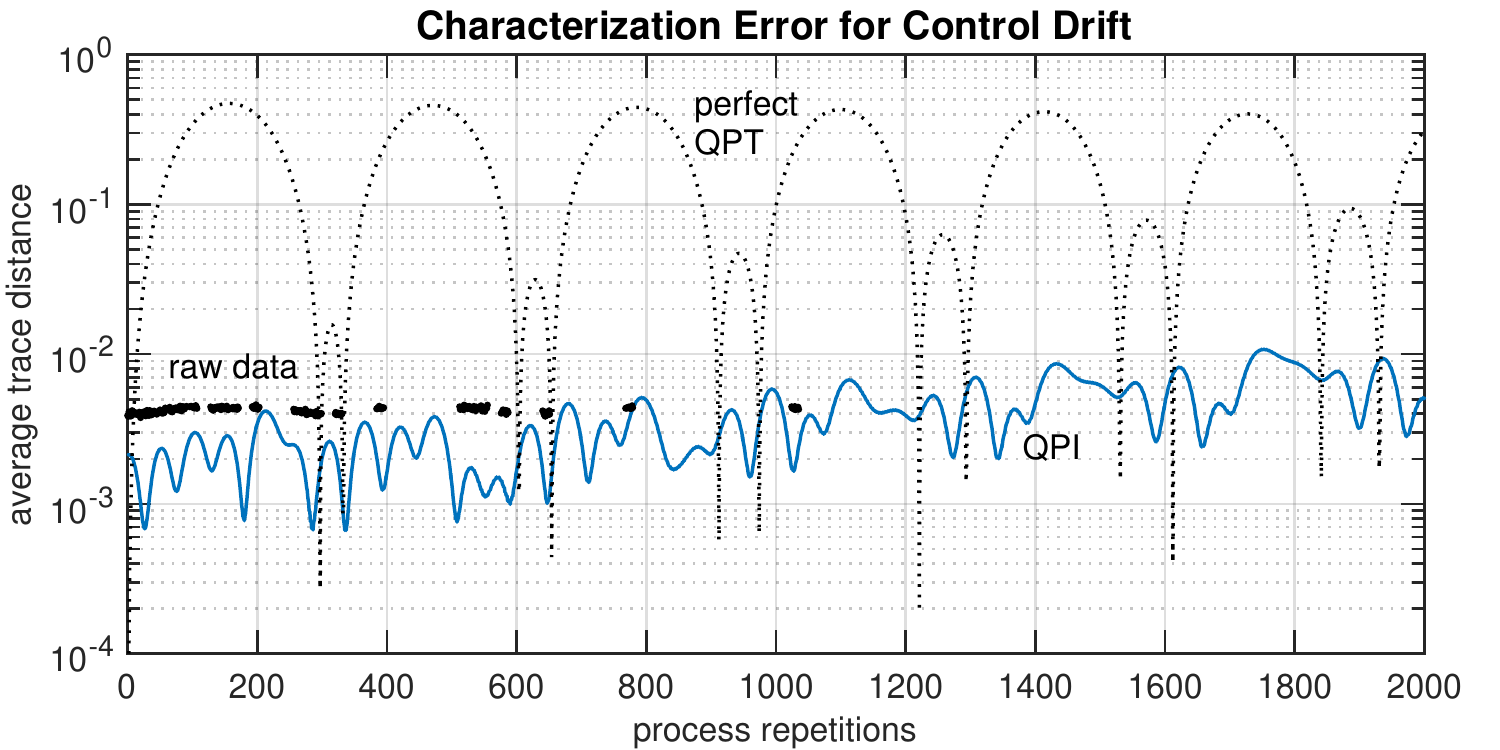}
\par\end{centering}
\caption{\label{fig: drift error}Average characterization error (solid line)
for a qubit undergoing $\pi$ rotations with small, sinusoidally varying
error in the rotation angle. Also shown for comparison are the raw
measurement error (dots) and error of perfect process tomography (dotted
line). }

\end{figure}

The evolution of the qubit under this process was simulated for two
different initial states, $\ket +_{z}$ and $\ket +_{x}$. For each
initial state, tomographic measurements of the qubit were simulated
at 304 times ranging from $t=0$ to $t=1035$, forming 57 flights
of length 12. Each measurement was repeated $10^{4}$ times. In nearly
all realizations, model inference yielded an 11-dimensional model.
Fig.~\ref{fig: drift error} shows the average characterization error
as a function of $t$.

\subsection{Coherent Qubit Leakage}

Leakage refers to the loss of a qubit, or more precisely, the transition
of a qubit to a non-computational state. This process is usually modeled
as irreversible and incoherent, in which case the qubit has no effect
on any subsequent operation. More realistically, qubit operations
create and manipulate superpositions of computational and non-computational
states \cite{Wood2017}. In such cases a qubit can effectively leave
and later return to the computational subspace, interfering with the
computational evolution. The particular details of the qubit's behavior
will depend strongly on the physics of the physical implementation
of the qubit and the control mechanisms employed. For this example
we use the 3-level anharmonic oscillator model which is applicable
to popular qubit technologies including superconducting flux qubits
and trapped ions \cite{Gambetta2011}. In this model, the control
pulse drives not only the intended $\ket 0\leftrightarrow\ket 1$
transition, but also off-resonant transitions between $\ket 1$ and
the next excited state $\ket 2$. The interaction Hamiltonian is
\begin{align}
H & =\Omega(t)\left(\ket 0\bra 1+\sqrt{2}\ket 1\bra 2+\text{H.c}\right)+\Delta\ket 2\bra 2
\end{align}
where $\Omega(t)$ is the control amplitude and $\Delta$ is the detuning
of the non-computational state. For simulations we used Gaussian control
pulses of width 0.25 time steps (truncated to have 1 time step duration)
and chose an excited state detuning $\Delta=20$ yielding a system
in which the leakage in and out of the computational space is small
but not negligible. For these parameter values the total probability
of state $\ket 2$ oscillates with a period of nearly 30 time steps,
with maximal value of a few percent. Meanwhile, the computational
state slowly precesses about the intended state.

\begin{figure}
\begin{centering}
\includegraphics{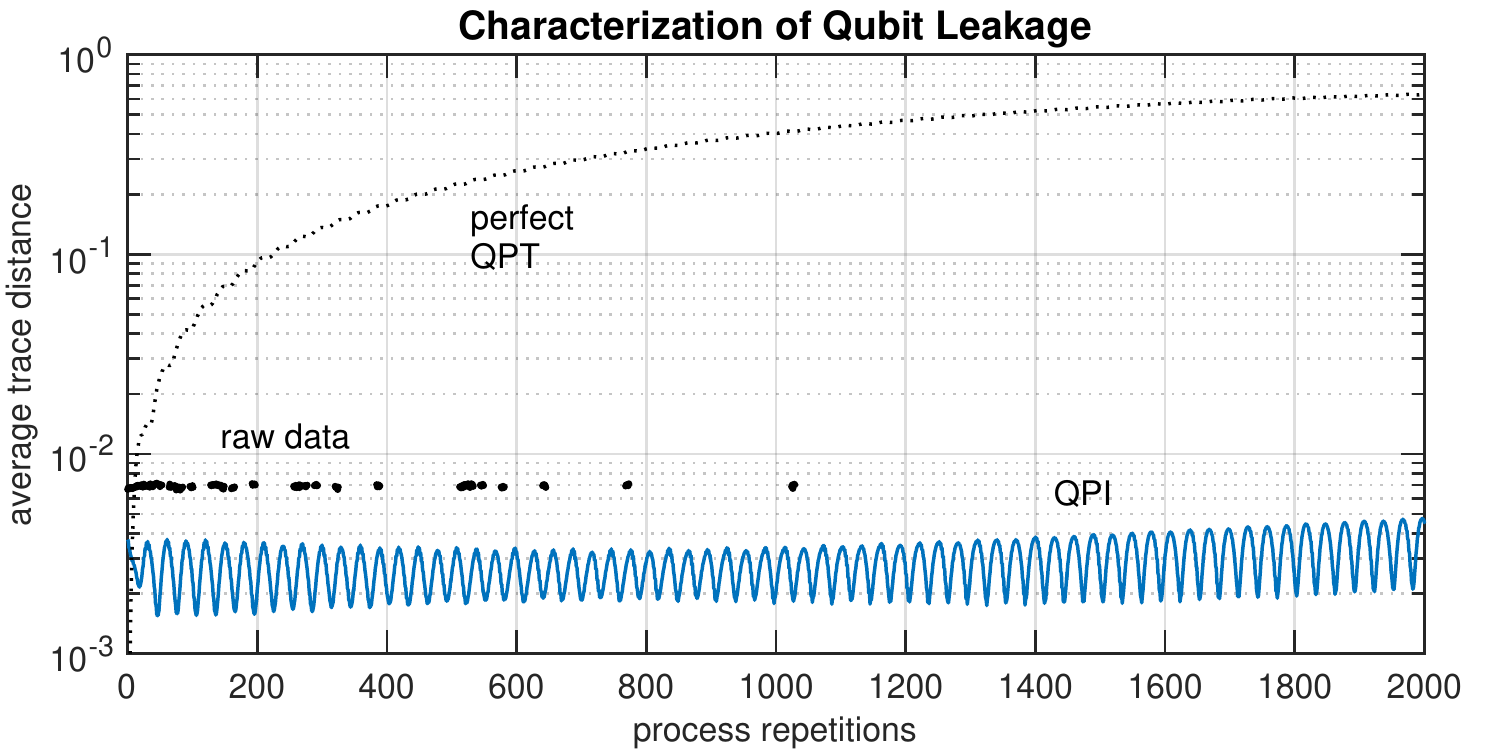}
\par\end{centering}
\caption{\label{fig: leak error}Average characterization error (solid line)
for a qubit undergoing bit flips in the anharmonic oscillator model.
Also shown for comparison are the raw measurement error (dots) and
error of perfect process tomography (dotted line). }

\end{figure}

The evolution of this 3-level system was simulated for four different
initial qubit states ($\ket +_{z}=\ket 0$, $\ket -_{z}=\ket 1$,
$\ket +_{x}$, and $\ket +_{y}$). For each initial state, tomographic
measurements of the qubit were simulated at 196 times ranging from
$t=0$ to $t=1029$, forming 57 flights of length 6. Each measurement
was repeated $10^{4}$ times. In nearly all cases, the model inference
yielded a 7-dimensional model. The average characterization error
for this process is shown in Fig.~\ref{fig: leak error}.

\subsection{A Qubit Coupled to a Single Impurity}

\begin{figure}
\begin{centering}
\includegraphics{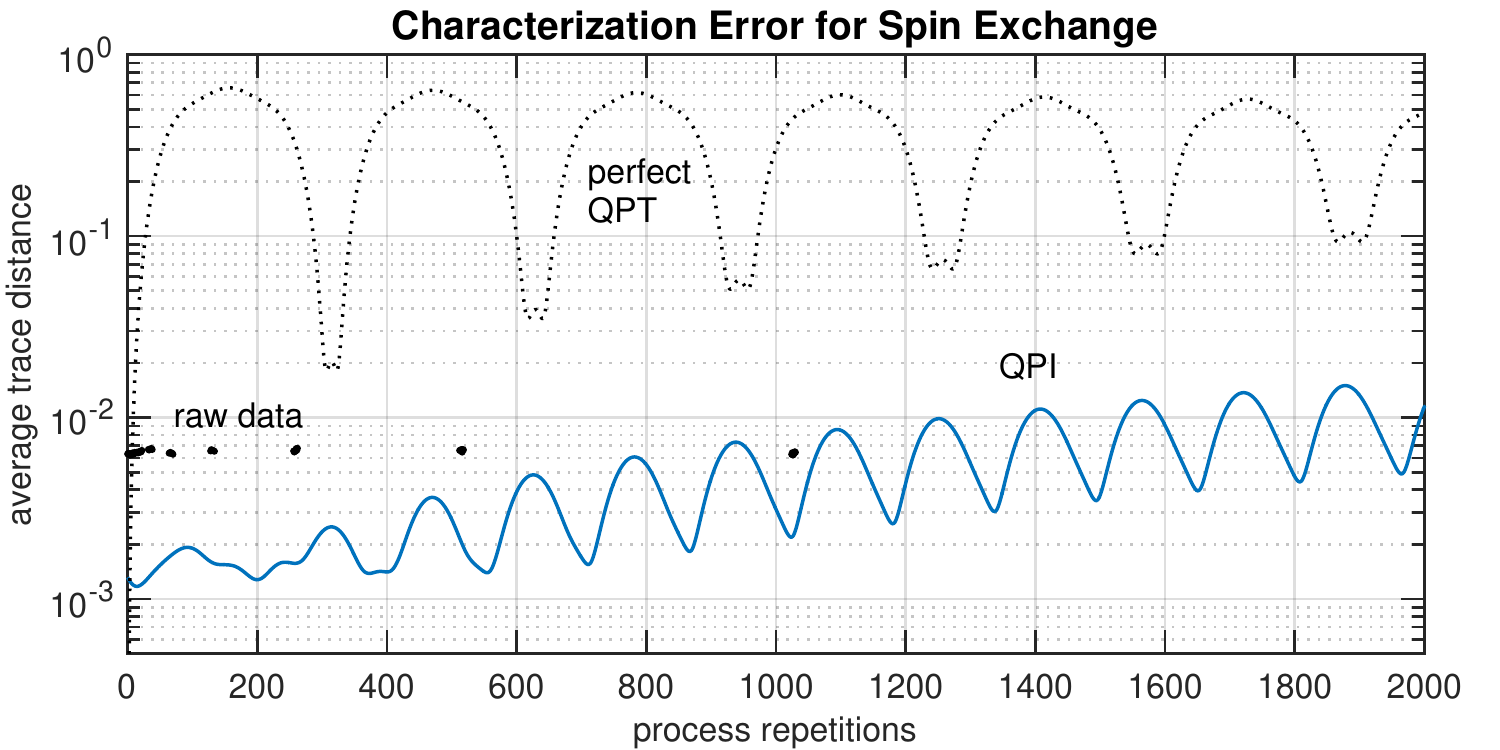}
\par\end{centering}
\caption{\label{fig: spin error}Average characterization error (solid line)
for a qubit weakly coupled to an impurity via a spin-exchange interaction.
Also shown for comparison are the raw measurement error (dots) and
error of perfect process tomography (dotted line). }

\end{figure}

Another potential source of error in qubit devices is unintended coupling
between qubits, or between a qubit and a nearby impurity. A simple
model for such coupling is the isotropic spin exchange Hamiltonian
\begin{align}
H & =\gamma\left(\sigma_{x}\otimes\sigma_{x}+\sigma_{y}\otimes\sigma_{y}+\sigma_{z}\otimes\sigma_{z}\right).
\end{align}
Taking $\gamma=0.01$ to model weak coupling, the evolution of a qubit
and impurity spin was simulated for three initial qubit states ($\ket +_{x}$,
$\ket +_{y}$, and $\ket +_{z}$). The impurity is always initially
in the state $\ket +_{z}$. For each initial state, tomographic measurements
were simulated at 64 times ranging from $t=0$ to $t=1030$, forming
12 flights of length 7. Each measurement was repeated $10^{4}$ times.
In nearly all cases, model inference yielded a 7-dimensional model.
This may be compared with the nominal dimension $4^{2}=16$ of a two
qubit system. The average characterization error for this process
is shown in Fig. \ref{fig: spin error}.

\subsection{$^{31}$P Qubit in Silicon}

\label{sec: 31P study}For our final case study, we considered a $^{31}\text{P}$
donor in $^{28}\text{Si}$ with $^{29}\text{Si}$ impurities. The
nuclear spin of the $^{31}\text{P}$ defines a qubit. The qubit is
manipulated by a time-varying magnetic field which also affects the
impurities. Meanwhile, the $^{31}\text{P}$ and $^{29}\text{Si}$
impurities interact via hyperfine coupling to the donor electron.
A detailed model for this system was formulated in \cite{Lougovski2017}.
We simulated the behavior of the qubit and impurities in response
to a sequence of pulses that each nominally produce a $\pi$ rotation
of the qubit (bit flip). For these simulations, the donor was imagined
to be at the center of a $(5\thinspace\text{nm})^{3}$ cube with a
$^{29}\text{Si}$ concentration of 800 ppm, which translates to 6
impurities in the qubit volume. The impurities were distributed randomly
throughout the volume; in all, 200 random configurations were simulated.
The background magnetic field $B_{z}$ was 1.5 T, the amplitude $B_{x}$
of the driving magnetic field was 0.15 T, and the ambient temperature
was 250 mK. (For further details see \cite{Lougovski2017}). 

\begin{figure}
\begin{centering}
\includegraphics{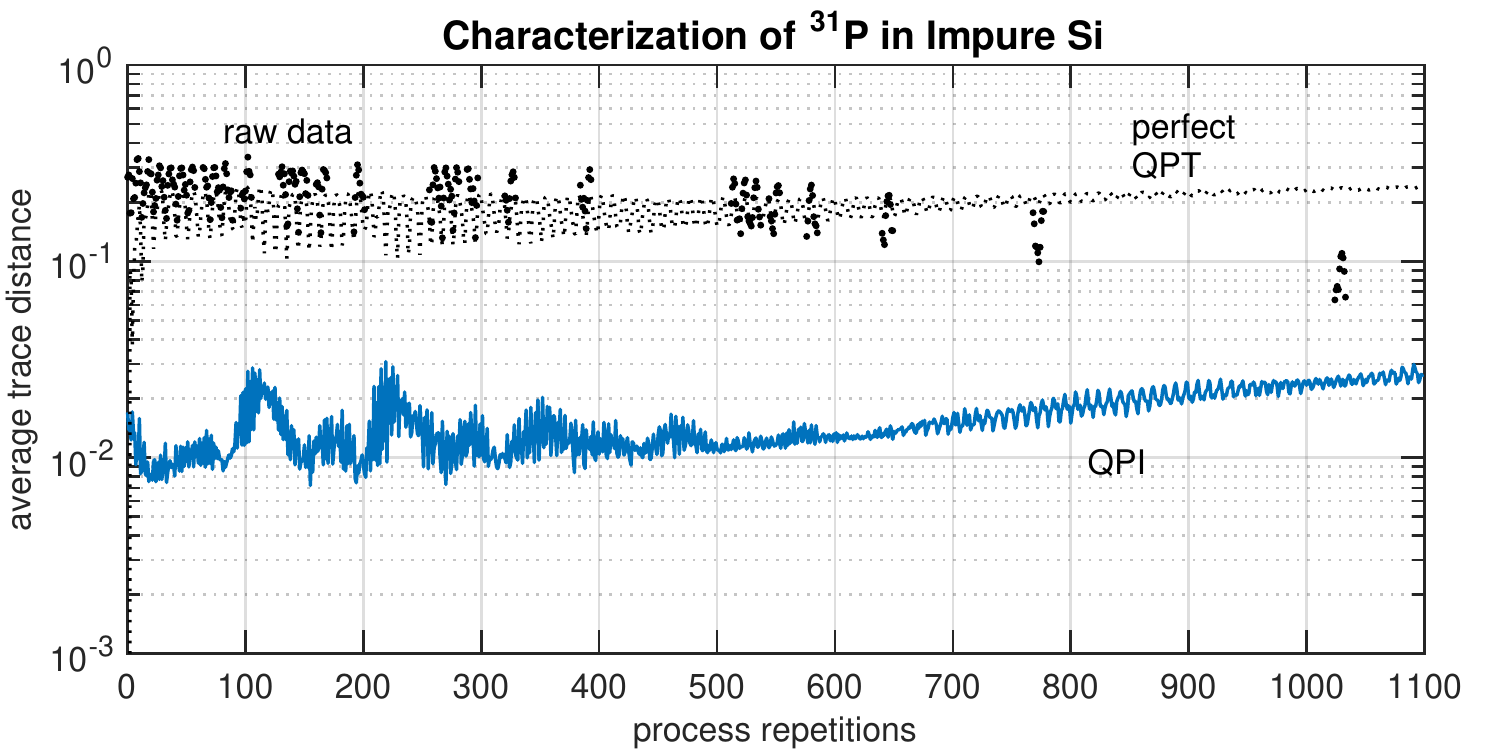}
\par\end{centering}
\caption{\label{fig: P-Si error}Average characterization error (solid line)
for a $^{31}$P donor qubit with 6 nearby $^{29}$Si impurities. Also
shown for comparison are the raw measurement error (dots) and error
of perfect process tomography (dotted line). }

\end{figure}

For each configuration of the impurities, the evolution of the donor
and impurities was simulated for 4 different initial qubit states
($\ket +_{x}$,$\ket +_{y}$,$\ket +_{z}$, and $\ket -_{z}$) and
with the impurities initially in thermal equilibrium. For each initial
state, tomographic measurements of the qubit were simulated at 272
different time steps ranging from $t=0$ to $t=1033$, forming 57
flights of length 10. Each measurement was repeated $10^{3}$ times.
In most realizations the inferred model dimension was 19 or 20, which
is much less than the nominal dimension of the $^{31}\mathrm{P}+^{29}\mathrm{Si}$
system (7 nuclear spins + 1 electron spin = 8 spins with a nominal
dimension of $(2^{8})^{2}=65,536$). Thus QPI reveals that the effective
interaction between the qubit and the impurities involves only a small
subspace of the available Hilbert space, an insight that is not readily
apparent from the underlying model. The average characterization error
for this system is shown in Fig.~\ref{fig: P-Si error}

\section{Discussion}

\label{sec: discussion}In all of the simulation studies above, our
QPI method was able to accurately capture and reproduce the observable
behavior of the qubit over very long time scales, with trace distance
errors on the order of $10^{-2}$. While this level of error would
not be impressive for tomography of the state at any one time, the
fact that this accuracy is maintained over more than a thousand time
steps implies that the process itself is actually characterized to
an accuracy on the order of $10^{-5}$. Notably, the inferred models
remained accurate well beyond the times at which measurements were
taken (although the errors did tend to slowly grow the farther in
time the models were extrapolated).

In contrast, standard quantum process tomography (which uses measurements
only at $t=0$ and $t=1$) yielded inaccurate predictions of qubit
behavior past a few tens of time steps, even under the assumption
of zero statistical measurement error. Arguably, it would be fairer
to compare QPI with a version of QPT that incorporates the data at
all times, but assumes the process is Markovian. This can be achieved
in our framework by restricting the model dimension to 4 for characterization
of a single qubit. In fact we attempted to perform such a comparison,
but found that dimension-limited (Markovian) models were such a poor
fit to the data that the model inference would not converge.

For the finite dimensional processes studied (leakage and spin exchange),
QPI inferred a process dimension that was equal to or close to the
true dimension. For the high-dimensional processes studied (control
drift and $^{31}\mathrm{P}+^{29}\mathrm{Si}$ systems), QPI was able
to reproduce the observed behavior using relatively low-dimensional
models. Fortunately, it was not necessary to use flights as long as
suggested by the worst-case bound of Lemma 1.

Although these simulation studies addressed processes nominally involving
a single qubit, QPI is in principle just as applicable to the characterization
of qudit or multiqubit operations. However, the state space size (and
hence also the cost of data collection and data analysis) grows exponentially
with the number of qubits. Thus like all forms of tomography, QPI
is really only feasible for characterizing small quantum systems.
Indeed, a possible concern is that QPI appears to require much more
experimental data than QPT, which already requires a moderately large
amount of data. To some extent this is to be expected: whereas QPT
needs only enough information to produce a 4-dimensional model for
a qubit process, QPI needs enough information to first of all determine
the effective size of the environment's memory and then determine
all the dynamical parameters involving the qubit, the pertinent part
of the environment, and their interaction. We expect, however, that
QPI can be made much more efficient than the studies above would suggest.
In these studies, all measurements were arbitrarily chosen to be performed
the same number of times. Almost certainly some measurements are more
informative than others. We expect that an adaptive testing strategy,
in which the current model estimate is used to choose the most informative
experiment to perform next, will significantly reduce the cost of
QPI \cite{Huszar2012,Mahler2013,Granade2015}.

As presented here, QPI has two main limitations. The first is that
it characterizes only one process. In the context of a quantum computer,
one would like a self-consistent characterization of all qubit operations
(``gates'') that may be performed; under the assumption of Markovianity,
GST provides just such a characterization. As formulated here, QPI
would yield a set of \emph{separate} characterizations, one for each
gate, that cannot be related. More specifically, the relationship
between latent variables in the models for different gates would be
indeterminate, i.e. one would not know whether different gates interact
with the same or different environmental degrees of freedom. We conjecture
that QPI can be extended to jointly characterize a set of processes:
The principles of the Ho-Kalman method extend in a straightforward
way to this case. What is unclear at present is how to extend Lemma
1, i.e.~how to identify a relatively small set of operation sequences
that are likely to yield a complete model. This is an obvious direction
for future work.

In principle, extending QPI to multiple processes would allow cross
talk and other spatial correlations to be characterized by treating
different combinations of simultaneous gates as different processes.
However, due to the very large number of different combinations this
would constitute (even in a small device), this does not seem to be
a practical approach to characterizing spatial correlations.

The other main limitation of QPI as described here is that it assumes
all measurements are final; no attempt is made to model the effect
of the measurement on the state of the system or environment. But
intermediate measurements are a key component of quantum error correction
protocols, thus the ability to characterize their impact on the remainder
of a computation is important. We suspect the ability to characterize
intermediate measurements would follow from the ability to characterize
multiple processes, since each outcome of a measurement can be treated
as a distinct, randomly selected process.

\section{Conclusion}

In conclusion, we have developed a new method for the characterization
of quantum dynamical processes, called quantum process identification.
QPI goes beyond state-of-the-art methods by providing the means to
systematically characterize non-Markovian dynamics as well as Markovian
dynamics. We presented a detailed experimental protocol and data analysis
procedure and demonstrated their effectiveness using numerical simulations
of realistic non-Markovian error processes affecting qubits. Directions
for future work include extending QPI to enable the characterization
of multiple processes and non-final measurements, and using adaptive
strategies to reduce its experimental cost. Apart from experimental
applications, the theory underlying QPI elucidates the relationship
between the temporal evolution of an open quantum system and its effective
environment, providing a useful alternative to existing representations
of non-Markovian quantum dynamics.

This research was sponsored by the Laboratory Directed Research and
Development Program of Oak Ridge National Laboratory, managed by UT-Battelle,
LLC, for the U. S. Department of Energy. The authors would like to
express appreciation to Nicholas Peters for helpful comments on the
preparation of this manuscript.

\bibliography{process_identification}

\appendix

\section{Proof of Lemma 1}

\label{sec: proof of Lemma}In this section we show that the minimum
number of time observations needed to guarantee complete characterization
of a discrete-time linear system depends on the number of latent (unmeasured)
degrees of freedom, not the total number of degrees of freedom. Let
$d$ be the true dimension of the system and let $a$ be the dimension
of the subspace spanned by the initial states and measured quantities.
We show that $\rank(H_{l+1})=d$ where $l=d-a$ and $H_{l}$ is the
block-Hankel matrix formed from $X(1),\ldots,X(2l-1)$. It follows
from the Ho-Kalman theory that $X(1),\ldots,X(2(l+1))$ are sufficient
to obtain a state model of the system .

The goal is to find matrices $L,M,R$ such that $X(k)=LM^{k-1}R$.
In the remainder of this section, $X(k)$ will be denoted more compactly
as $X_{k}$. For convenience, we assume that redundant initial states
and redundant measurements have been omitted, so that $X(k)$ is $a\times a$.
Since $\rank X_{i}\le a$ for every $X_{i}$, we may choose a representation
in which $L$ has the form $L=\left[\begin{array}{cc}
1_{a\times a} & 0_{a\times l}\end{array}\right]$. It follows that $R=\left[\begin{array}{c}
X_{0}\\
Y_{0}
\end{array}\right]$ where $Y_{0}$ is some unknown $l\times a$ matrix. We write $M$
as
\begin{align}
M & \equiv\left[\begin{array}{cc}
A & B\\
C & D
\end{array}\right]
\end{align}
where $A$ is $a\times a$, $D$ is $l\times l$, and $B$ and $C$
are sized correspondingly. We have
\begin{align}
\left[\begin{array}{c}
X_{n}\\
Y_{n}
\end{array}\right] & \equiv M^{n}\left[\begin{array}{c}
X_{0}\\
Y_{0}
\end{array}\right]\\
 & =\left[\begin{array}{cc}
A & B\\
C & D
\end{array}\right]\left[\begin{array}{c}
X_{n-1}\\
Y_{n-1}
\end{array}\right].
\end{align}
Then
\begin{align}
X_{n+1} & =AX_{n}+BY_{n}\\
 & =AX_{n}+B(CX_{n-1}+DY_{n-1})\\
 & \vdots\\
 & =AX_{n}+\left(\sum_{i=1}^{k}BD^{i-1}CX_{n-i}\right)+BD^{k}Y_{n-k}
\end{align}
or 
\begin{gather}
X_{n+k+1}=AX_{n+k}+\left(\sum_{i=1}^{k}BD^{i-1}CX_{n+l-i}\right)+BD^{k}Y_{n}\label{eq: X Y recurrence 1}
\end{gather}
for $k=0,1,2,\ldots,n$. Let $k=l$. Our goal is to eliminate $Y_{n}$
and obtain a recurrence relation involving only the $X_{j}$'s. By
the Cayley-Hamilton theorem there are coefficients $\alpha_{0},\ldots,\alpha_{l-1}$
such that $D^{l}=\sum_{j=0}^{l-1}\alpha_{i}D^{i}$. Thus
\begin{gather}
X_{n+l+1}=AX_{n+l}+\left(\sum_{i=1}^{l}BD^{i-1}CX_{n+l-i}\right)+\sum_{j=0}^{l-1}\alpha_{j}BD^{j}Y_{n}.\label{eq: X Y recurrence 2}
\end{gather}
Setting $k$ to $j$ in (\ref{eq: X Y recurrence 1}) yields
\begin{align}
X_{n+j+1} & =AX_{n+j}+\left(\sum_{i=1}^{j}BD^{i-1}CX_{n+j-i}\right)+BD^{j}Y_{n}.
\end{align}
Solving for $BD^{j}Y_{n}$ and substituting into (\ref{eq: X Y recurrence 2})
gives
\begin{align}
X_{n+l+1} & =AX_{n+l}+\left(\sum_{i=1}^{l}BD^{i-1}CX_{n+l-i}\right)+\sum_{j=0}^{l-1}\alpha_{j}\left(X_{n+j+1}-AX_{n+j}-\sum_{i=1}^{j}BD^{i-1}CX_{n+l-i}\right)\label{eq: X recurrence}\\
 & =AX_{n+l}+\sum_{i=1}^{l}\left(1-\sum_{j=i}^{l-1}\alpha_{j}\right)BD^{i-1}CX_{n+l-i}+\sum_{j=0}^{l-1}\alpha_{j}\left(X_{n+j+1}-AX_{n+j}\right)
\end{align}
The important thing to note about this last equation is that $X_{n+l+1}$
is written in terms of $X_{n},\ldots,X_{n+l}$. More precisely, there
are matrices $Q_{0},\ldots,Q_{l}$ such that
\begin{gather}
X_{n+l+1}=\sum_{i=0}^{l}Q_{i}X_{n+i}.
\end{gather}
In other words, the rows of $X_{n+l+1}$ are linear combinations of
the rows of $X_{n},\ldots,X_{n+l}$. (Note that if the $Q_{i}$'s
were scalars instead of matrices, the effective dimension of the system
would be at most $l+1$.) By extension, 
\begin{align}
\mathcal{T}_{n+l+1} & =\sum_{i=0}^{l}Q_{i}\mathcal{T}_{n+i}
\end{align}
where $\mathcal{T}_{k}\equiv\left[\begin{array}{ccc}
X_{k} & X_{k+1} & \cdots\end{array}\right]$ is the $k$th block of rows of $H_{\infty}$. Thus $\mathcal{T}_{1},\ldots,\mathcal{T}_{l+1}$
span $\mathcal{T}_{l+1}$ and, by induction, all rows of $H_{\infty}$.

Now, we could just as well have chosen a representation in which $R=[\begin{array}{cc}
1_{a\times a}; & 0\end{array}]$ and $L=\left[\begin{array}{cc}
X_{0} & Z_{0}\end{array}\right]$, and written 
\begin{gather}
\left[\begin{array}{c}
X_{n}\\
Z_{n}
\end{array}\right]=\left[\begin{array}{c}
X_{0}\\
Z_{0}
\end{array}\right]M^{n}=\left[\begin{array}{c}
X_{n-1}\\
Z_{n-1}
\end{array}\right]\left[\begin{array}{cc}
A & B\\
C & D
\end{array}\right].
\end{gather}
This is the same problem as above, except that $Y_{n}$ is replaced
by $Z_{n}$, $B$ and $C$ swap roles, and recurrence relations involve
factors of $A,B,C,D$ on the right rather than on the left. A derivation
analogous to that above would show that there exist matrices $Q_{0}^{\prime},\ldots,Q_{l}^{\prime}$
such that 
\begin{align}
X_{n+l+1} & =\sum_{i=0}^{l}X_{n+i}Q_{i}^{\prime}
\end{align}
by which we conclude that the first $a(l+1)$ columns span all the
columns of $H_{\infty}$. Thus $\rank(H_{l+1})=\rank(H_{\infty})=a+l$.
It follows from the Ho-Kalman theory that $X_{1},\ldots,X_{2(l+1)}$
are sufficient to determine any system of dimension up to $a+l$.

Conversely, observing the output at less than $2(l+1)$ times is generally
insufficient to reproduce the system's behavior. We omit a detailed
proof, but note that this fact can be established by constructing
a $d$-dimensional system with a size $l$ shift register that does
not reveal its presence until the $(l+1)$th time step. For this process
$\rank H_{l}<\rank H_{l+1}=d$. Similarly, it is easy to show that
observing the output at non-consecutive times is also generally insufficient
to construct a correct model. Consider a scalar system whose impulse
response function (starting at time $t=0$) cycles through the values
$\alpha\to\beta\to\alpha\to\gamma\to\ldots$. If the outputs are observed
only at $t=0,1,2,4,\ldots,2^{k},\ldots,$ the value $\gamma$ would
never be observed and a two-state system with output $\alpha\rightarrow\beta\rightarrow\alpha\rightarrow\beta\rightarrow\cdots$
would be inferred.

\section{Dimension Estimation}

\label{sec: dimension estimation} In this section we consider the
problem of estimating the dimension $d$ of a process from an experimental
estimate $\tilde{H}$ of a Ho-Kalman matrix $H$. Ho and Kalman showed
if $H$ is sufficiently large, then $d$ is just the rank of $H$.
But $\tilde{H}$, which is a random perturbation of $H$, has a rank
that is generally much higher than $d$. The singular values of $\tilde{H}$
are much more informative than the rank, as they reveal the magnitude
of each dimension's contribution to the data. If the statistical
errors in the data are small, all but $d$ singular values of $\tilde{H}$
will be small. Here we derive a reliable criterion for determining
which singular values are small enough to be considered spurious.

A naive approach would be to set a threshold for each singular value\textemdash say,
based on its estimated variance\textemdash and count the number of
singular values that exceed their threshold. There are problems with
this approach, however, the most serious being that the number of
singular values that exceed their threshold simply by chance grows
with the size of $\tilde{H}$, i.e. grows with the number of experiments
performed. A way to overcome this problem is to test singular values
collectively rather than individually: Given a threshold for statistical
significance of the $k$ smallest singular values, one finds the largest
$k$ for which the threshold is not met and takes $d$ to be the number
of singular values that remain.

An appropriate threshold can be derived from the uncertainty in the
raw data. Each quantity $\tilde{F}_{i,m}^{(t)}$ that appears in $\tilde{H}$
is an independent random variable, distributed binomially with mean
$F_{i,m}^{(t)}$ and variance $F_{i,m}^{(t)}(1-F_{i,m}^{(t)})/N_{i,m}^{(t)}$.
Let $\epsilon_{i,t,m}=\tilde{F}_{i,m}^{(t)}-F_{i,m}^{(t)}$ denote
the statistical deviation of $\tilde{F}_{i,m}^{(t)}$ and let $\partial_{(i,t,m)}H\equiv\partial H/\partial F_{i,m}^{(t)}$.
Then
\begin{align}
\tilde{H} & =H+\sum_{x}(\partial_{x}H)\epsilon_{x}
\end{align}
where $x$ ranges over all experiments $(i,t,m)$. Let $h_{1}\times h_{2}$
be the size of $H$ and let $h=\min(h_{1},h_{2})$. Let $H=USV^{\dagger}$
be a ``full'' singular value decomposition of $H$, i.e. $U$ is
$h_{1}\times h_{1}$ and $V$ is $h_{2}\times h_{2}$, and $S$ is
$h_{1}\times h_{2}$ with diagonal elements $s_{1}\ge s_{2}\ge\cdots\ge s_{h}$.
Let $U_{k}$ be the matrix consisting of columns $k$ through $h_{1}$
of $U$, and $V_{k}$ the matrix consisting of columns $k$ through
$h_{2}$ of $V$. 

 Suppose $H$ has rank $r$. Then $s_{1},\ldots,s_{r}\ne0$ and $s_{r+1}=\cdots=s_{h}=0$.
According to perturbation theory, the corresponding singular values
$\tilde{s}_{r+1},\ldots,\tilde{s}_{h}$ of $\tilde{H}$ are the singular
values of 
\begin{align}
A & \equiv U_{r+1}^{\dagger}\tilde{H}V_{r+1}\\
 & =U_{r+1}^{\dagger}\left(H+\sum_{x}(\partial_{x}H)\epsilon_{x}\right)V_{r+1}.
\end{align}
Since $U_{r+1}^{\dagger}HV_{r+1}=0$ we have
\begin{align}
A & =\sum_{x}A_{x}\epsilon_{x}
\end{align}
where
\begin{align}
A_{x} & \equiv U_{r+1}^{\dagger}(\partial_{x}H)V_{r+1}.
\end{align}
Let $\chi_{r}$ denote the residual ``energy'' of the $h-r$ smallest
singular values,
\begin{align}
\chi_{r} & =\sum_{i>r}\tilde{s}_{i}^{2}\\
 & =\Tr(A^{T}A)\\
 & =\sum_{x,y}\Tr(A_{x}^{T}A_{y})\epsilon_{x}\epsilon_{y}.
\end{align}
Owing to the independence of experiments, we have $\left<\epsilon_{x}\epsilon_{y}\right>=0$
for $x\ne y$, yielding
\begin{align*}
\left<\chi_{r}\right> & =\sum_{i,j>r}\sum_{x}B_{ij}^{(x)}
\end{align*}
where $B^{(x)}\equiv A_{x}\sqrt{\left<\epsilon_{x}^{2}\right>}$.

If the threshold for rejecting $\tilde{s}_{r},\ldots,\tilde{s}_{h}$
were set at $\left<\chi_{r}\right>$, then approximately half of the
time the subset $\tilde{s}_{d+1},\ldots,\tilde{s}_{h}$ would be incorrectly
accepted as statistically significant, resulting in an overestimate
of $d$. A higher threshold reduces the probability of this occurring,
but increases the probability of rejecting $\tilde{s}_{d},\ldots,\tilde{s}_{h}$
and thereby underestimating $d$. Thus the threshold should be just
high enough to reject $\tilde{s}_{d+1},\ldots,\tilde{s}_{h}$ a majority
of the time. To that end we calculate the variance of $\chi_{r}$.
We have 
\begin{align*}
\chi_{r}^{2} & =\Tr(A^{T}A)^{2}\\
 & =\sum_{x,y}\sum_{p,q}\Tr(A_{x}^{T}A_{y})\Tr(A_{p}^{T}A_{q})\epsilon_{x}\epsilon_{y}\epsilon_{p}\epsilon_{q}.
\end{align*}
Since the deviations are independent and zero-mean, $\left<\epsilon_{x}\epsilon_{y}\epsilon_{p}\epsilon_{q}\right>$
vanishes unless it is of the form $\left<\epsilon_{x}^{4}\right>$
or $\left<\epsilon_{x}^{2}\epsilon_{y}^{2}\right>$. The second form
simplifies to $\left<\epsilon_{x}^{2}\right>\left<\epsilon_{y}^{2}\right>$.
Provided the statistical errors are small compared to the quantities
being estimated, $\epsilon_{x}$ has an approximately normal distribution,
for which $\left<\epsilon_{x}^{4}\right>=3\left<\epsilon_{x}^{2}\right>^{2}$.
Then
\begin{align}
\left<\chi_{r}^{2}\right> & =3\sum_{x}\Tr(A_{x}^{T}A_{x})^{2}\left<\epsilon_{x}^{2}\right>^{2}+\sum_{x}\sum_{y\ne x}\left(\Tr(A_{x}^{T}A_{x})\Tr(A_{y}^{T}A_{y})+2\Tr(A_{x}^{T}A_{y})^{2}\right)\left<\epsilon_{x}^{2}\right>\left<\epsilon_{y}^{2}\right>\nonumber \\
 & =\left(\sum_{x}\Tr(A_{x}^{T}A_{x})\left<\epsilon_{x}^{2}\right>\right)^{2}+2\sum_{x,y}\Tr(A_{x}^{T}A_{y})^{2}\left<\epsilon_{x}^{2}\right>\left<\epsilon_{y}^{2}\right>\nonumber \\
 & =\left<\chi_{r}\right>^{2}+2\sum_{i,j>r}\sum_{k,l>r}\left(\sum_{x}B_{ij}^{(x)}B_{kl}^{(x)}\right)^{2}
\end{align}
where
\begin{align}
D_{i,j;k,l} & \equiv\sum_{x}B_{ij}^{(x)}B_{kl}^{(x)}.
\end{align}

Our criterion for accepting the hypothesis $s_{r+1}=s_{r+2}=\cdots=0$
is
\begin{align}
\chi_{r} & \le\left<\chi_{r}\right>+\sqrt{\Var\chi_{r}}\\
 & =\sum_{i,j>r}\sum_{x}B_{i,j}^{(x)}+\sqrt{2\sum_{i,j>r}\sum_{k,l>r}\left(\sum_{x}B_{ij}^{(x)}B_{kl}^{(x)}\right)^{2}}.\label{eq: singular value criterion}
\end{align}
We take $d$ be the smallest $r$ for which this criterion is met.
In theory the probability of overestimating $d$ is then at most 0.16.

In practice this criterion cannot be applied exactly: The quantities
on the right side of eq.~(\ref{eq: singular value criterion}) are
not known, but can only be estimated. The singular values may not
even be in quite the right order due to statistical variations. Furthermore,
the approximations made in the derivation might not always be accurate.
Nevertheless, in our simulations we observed that the criterion (\ref{eq: singular value criterion})
is effective: It usually yields an estimate of $d$ quite close to
the true value, provided one has taken enough data.

\section{Progressive Fitting of Data Blocks}

\label{sec: improved estimate}For large times $t$, the experimental
quantity $F^{(t)}=ST^{t}P$ is an extremely nonlinear function of
$T$, presenting a considerable challenge for estimation of the parameters
$S,T,P$ from the data. Building on the approach described in \cite{Blume-Kohout2015},
we devised a progressive fitting method which first uses low-$t$
data to obtain a coarse estimate of $T$, then gradually incorporates
data with higher $t$ to increase the accuracy of the model. Our implementation
involves multiple passes over the data, alternating between optimization
of $T$ given the current estimate of $S,P$ and optimization of $S,P$
given the current estimate of $T$. We found that if too much or too
little data was used at any given step, either $(S,P)$ or $T$ could
``lock in'' to the wrong region of parameter space from which the
search could not recover. The procedure below was carefully honed
to ensure that at each step, only reliable intermediate estimates
were used. In our studies, 5-15 passes were usually sufficient to
obtain good fits to the data.

Let $H^{(b)}$ denote the $b$th block of data (Fig.~\ref{fig: data structures}c),
$W^{(b)}$ the matrix of corresponding statistical weights, and $N_{b}$
the number of elements in these blocks. (The weights $W^{(b)}$ are
adjusted to account for the different multiplicity of different experiments,
so that each experiment has the same total weight.) The function to
be minimized is
\begin{align}
\Phi_{b}(A,T,B) & \equiv\frac{1}{N_{b}}\sum_{b^{\prime}=0}^{b}\sum_{i\in\mathcal{I}}\sum_{m\in\mathcal{M}}W_{i,m}^{(b^{\prime})}(AT^{\varrho_{b^{\prime}}}B-H^{(b^{\prime})})_{i,m}^{2}\label{eq: block fitting function}
\end{align}
which is the average weighted error of the model over blocks $0$
through $b$. Under the true model, the distribution of $\Phi_{b}$
has a strong peak at 1. A value significantly larger than this (say,
1.5) indicates a poor fit. Model fitting with the correct dimension
usually yields $\Phi_{b}$ values around 0.8; overfitting leads to
$\Phi_{b}$ around 0.5.

Recall that $\bar{b}$ is the index of the last block. Let $b_{\text{highest}}$
denote the index of the highest block for which a fit has been attempted.
The fitting procedure is as follows:
\begin{enumerate}[label=3.\arabic{enumi}]
\item Let $(L,T,R)$ be the outputs of Ho-Kalman estimation. Set $A$ equal
to $L$ and set $B$ equal to the first $n(l+1)$ columns of $R$.
Find the smallest $b$ such that $\Phi_{b}>1.5$ and set $b_{\text{highest}}$
to this value (or to $\bar{b}$ if there is no such $b$.)
\item Keeping $T$ fixed, optimize $A,B$. This is accomplished by minimizing
a modified form of eq.\,(\ref{eq: block fitting function}) with
$b=\bar{b}$. Whereas the current estimate of $T$ is used for blocks
$b^{\prime}\le b_{\text{highest}}$, for blocks $b^{\prime}>b_{\text{highest}}$
we replace $T^{\varrho_{b^{\prime}}}$ by the matrix $Y^{(b^{\prime})}$
that is optimal with respect to $A,B$. In this way $A,B$ are made
to fit all the data but are not constrained by high powers of $T$
that have not yet been determined to be accurate. Minimization can
be accomplished via iterative linear algebraic methods.
\item Keeping $A,B$ as fixed, optimize $T$.
\begin{enumerate}[label=3.\arabic{enumi}.\arabic{enumii}]
\item Set $b=2$.
\item If $b<b_{\text{highest}}$ and $\Phi_{b}\le1.5$, continue to 3.3.3.
Otherwise, find $T$ that minimizes $\Phi_{b}$, keeping $A,B$ fixed.
This can be accomplished by various nonlinear optimization methods;
we found the Gauss-Newton method with line search to work well.
\item If $b<\bar{b}$ and $\Phi_{b}\le1.5$, increase $b$ by 1, set $b_{\text{highest }}=\max(b,b_{\text{highest}})$,
and go back to 3.3.2.
\end{enumerate}
\item If $b=\bar{b}$ and $\Phi_{\bar{b}}$ did not improve significantly
(say, by more than 0.001) compared to the previous pass, optimization
has converged; continue to step 3.5. Otherwise, go back to step 3.2
(perform another pass).
\item If $\Phi_{\bar{b}}\le1.5$, the fit is deemed a success; in this case
return $T$, return the first $m$ rows of $A$ for $S$, and return
the first $n$ columns of $B$ for $P$. Otherwise, increase $d$
by one, perform Ho-Kalman estimation again (stage 2, described in
Section \ref{sec: model inference}), and start again at step 3.1.
\end{enumerate}

\section{Final Model Optimization}

\label{sec: final estimate}The fourth and final stage of our model
inference procedure is to fit the model $F^{(t)}=ST^{t}P$ directly
to the data. This last stage yields the most accurate models but requires
a very good starting estimate $S,T,P$. As in the previous stage,
we minimize a sum of weighted squared errors. But unlike in previous
stages, the weights are determined by the model itself rather than
estimated from the data, making the total squared error a good proxy
for the negative log likelihood. Besides being simpler to work with
than the log likelihood, weighted errors allow out-of-range probabilities
(less than 0 or grater than 1) to be handled much more gracefully.
In this stage we also impose soft constraints on the eigenvalues $\{\lambda_{i}\}$
of $T$: Since the process in question is presumed to occur in a finite
state space, the eigenvalues of $T$ should not have magnitude greater
than 1, as otherwise repeated application of the process would eventually
take the state out of the state space. 

The function to be minimized is
\begin{align}
\Psi(S,T,P) & \equiv\frac{1}{\left|\mathcal{I}\right|\left|\mathcal{M}\right|\left|\mathcal{T}\right|}\sum_{t\in\mathcal{T}}\sum_{i\in\mathcal{I}}\sum_{m\in\mathcal{M}}W_{i,m}^{(t)}\left(S,T,P\right)\left(F^{(t)}-\tilde{F}^{(t)}\right)_{i,m}^{2}+E(T).\label{eq: final optimization function}
\end{align}
where $E(T)=\sum_{i}\max(0,\left|\lambda_{i}\right|-1)^{2}$ is the
eigenvalue constraint. Nominally, $W_{i,m}^{(t)}$ is the inverse
variance of $F_{i,m}^{(t)}$. But since the inverse variance becomes
negative or infinite if the model ever predicts $F_{i,m}^{(t)}\le0$
or $F_{i,m}^{(t)}\ge1$, we take
\begin{align}
W_{i,m}^{(t)} & =\frac{1}{V_{i,m}^{(t)}+\sqrt{V_{i,m}^{(t)2}+4\beta_{i,m}^{(t)2}}}
\end{align}
where
\begin{align}
V_{i,m}^{(t)} & =\frac{F_{i,m}^{(t)}(1-F_{i,m}^{(t)})}{N_{i,m}^{(t)}}
\end{align}
is the theoretical variance of $F_{i,m}^{(t)}$ and $\beta_{i,m}^{(t)}$
is a small buffer term that ensures $W_{i,m}^{(t)}$ remains finite,
continuous, and positive. We start with $\beta_{i,m}^{(t)}=1/N_{i,m}^{(t)}$.
We then proceed to minimize \ref{eq: final optimization function}
using the Gauss-Newton method with line search. After each search
step the validity of the model predictions are assessed. If any $F_{i,m}^{(t)}$
is not in the interval $[0,1]$, $\beta_{i,m}^{(t)}$ is reduced by
5\%. This increases $W_{i,m}^{(t)}$ outside $[0,1]$ and near its
endpoints, thereby more strongly encouraging $F_{i,m}^{(t)}$ to become
valid. If after a search step all the predicted probabilities are
valid and $\Psi$ was not significantly improved, the search concludes.
\end{document}